\documentclass[twocolumn,pre]{revtex4}

\usepackage{dcolumn}
\usepackage{amsmath}

\usepackage{graphicx}
\usepackage{rotating}

\begin{document}

\newcommand{\defn}{\textbf}
\newcommand{\half}{\mbox{$\frac12$}}
\renewcommand{\d}{{\rm d}}
\renewcommand{\i}{{\rm i}}
\renewcommand{\O}{{\rm O}}
\newcommand{\e}{{\rm e}}
\newcommand{\set}[1]{\lbrace#1\rbrace}
\newcommand{\av}[1]{\langle#1\rangle}
\newcommand{\eref}[1]{(\ref{#1})}
\newcommand{\etal}{{\it{}et~al.}}
\newcommand{\vG}{\mathbf{G}}
\newcommand{\vf}{\mathbf{f}}
\newcommand{\vp}{\mathbf{p}}
\newcommand{\vq}{\mathbf{q}}
\newcommand{\vA}{\mathbf{A}}
\newcommand{\vI}{\mathbf{I}}
\newcommand{\cN}{\mathcal{N}}
\newcommand{\cov}{\mathop{\rm cov}}
\newcommand{\var}{\mathop{\rm var}}

\newlength{\figurewidth}
\ifdim\columnwidth<10.5cm
  \setlength{\figurewidth}{0.95\columnwidth}
\else
  \setlength{\figurewidth}{10cm}
\fi
\setlength{\parskip}{0pt}
\setlength{\tabcolsep}{6pt}

\title{Assortative mixing in networks}
\author{M. E. J. Newman}
\affiliation{Department of Physics, University of Michigan, Ann Arbor,
MI 48109--1120}
\affiliation{Santa Fe Institute, 1399 Hyde Park Road, Santa Fe, NM 87501}
\begin{abstract}
  A network is said to show assortative mixing if the nodes in the network
  that have many connections tend to be connected to other nodes with many
  connections.  We define a measure of assortative mixing for networks and
  use it to show that social networks are often assortatively mixed, but
  that technological and biological networks tend to be disassortative.  We
  propose a model of an assortative network, which we study both
  analytically and numerically.  Within the framework of this model we find
  that assortative networks tend to percolate more easily than their
  disassortative counterparts and that they are also more robust to vertex
  removal.
\end{abstract}
\maketitle

Many systems take the form of networks---sets of vertices joined together
by edges---including social networks, computer networks, and biological
networks~\cite{Strogatz01,AB02,DM02}.  A variety of models of networks have
been proposed and studied in the physics literature, many of which have
been successful at reproducing features of networks in the real
world~\cite{WS98,NSW01,BA99b}.  One particularly well-studied model is the
cumulative advantage or preferential attachment
model~\cite{Simon55,Price76,BA99b,KRL00,DMS00} in which the probability of
a given source vertex forming a connection to a target vertex is some
(usually increasing) function of the degree of the target vertex.  (The
degree of a vertex is the number of other vertices to which it is
attached.)  Preferential attachment processes are widely accepted as the
probable explanation for the power-law and other skewed degree
distributions seen in many networks~\cite{Price65,AJB99,FFF99,ASBS00}.

However, there is an important element missing from these as well as other
network models: in none of these models does the probability of attachment
to the target vertex depend also on the degree of the
\emph{source} vertex.  In the real world on the other hand such
dependencies are common.  Many networks show ``assortative mixing'' on
their degrees, i.e.,~a preference for high-degree vertices to attach to
other high-degree vertices.  Others show disassortative
mixing---high-degree vertices attach to low-degree ones.  In this paper we
first demonstrate the presence of assortative mixing in a variety of
networks by direct measurement, and then argue, using exactly solvable
models and numerical simulations, that assortative mixing can have a
substantial effect on the behavior of networked systems.  Models that do
not take it into account will necessarily fail to reproduce correctly many
of the behaviors of real-world networked systems.

Consider then a network, represented in the simplest case by an undirected
graph of $N$ vertices and $M$ edges, with degree distribution~$p_k$.  That
is, $p_k$~is the probability that a randomly chosen vertex on the graph
will have degree~$k$.  Now consider a vertex reached by following a
randomly chosen edge on the graph.  The degree of this vertex is not
distributed according to~$p_k$.  Instead it is biased in favor of vertices
of high degree, since more edges end at a high-degree vertex than at a
low-degree one.  This means that the degree distribution for the vertex at
the end of a randomly chosen edge is proportional~$kp_k$, rather than
just~$p_k$.  In this paper, we will usually be interested not in the total
degree of such a vertex, but in the
\emph{remaining degree}---the number of edges leaving the vertex other than
the one we arrived along.  This number is one less than the total degree
and hence is distributed in proportion to~$(k+1)p_{k+1}$.  The correctly
normalized distribution~$q_k$ of the remaining degree is then
\begin{equation}
q_k = {(k+1) p_{k+1}\over\sum_j j p_j}.
\label{defsqk}
\end{equation}

Following Callaway~\etal~\cite{Callaway01}, we now define the
quantity~$e_{jk}$ to be the joint probability distribution of the remaining
degrees of the two vertices at either end of a randomly chosen
edge~\footnote{A related quantity has been studied by Krapivsky and
  Redner~\cite{KR01} in the context of the model of Barab\'asi and
  Albert~\cite{BA99b}.  That quantity however, which is denoted~$n_{kl}$,
  is more complex than the one used here, being asymmetric in its indices,
  because one index is designated as being the ``ancestral'' index with
  respect to the order in which the graph was grown.}.  This quantity is
symmetric in its indices on an undirected graph $e_{jk}=e_{kj}$, and obeys
the sum rules
\begin{equation}
\sum_{jk} e_{jk} = 1,\qquad \sum_j e_{jk} = q_k.
\label{sumrules}
\end{equation}

In a network with no assortative (or disassortative) mixing $e_{jk}$ takes
the value~$q_j q_k$.  If there is assortative mixing, $e_{jk}$~will differ
from this value and the amount of assortative mixing can be quantified by
the connected degree-degree correlation function $\av{jk} - \av{j}\av{k} =
\sum_{jk} jk (e_{jk}-q_j q_k)$, where $\av{\ldots}$ indicates an average
over edges~\cite{Callaway01}.  This correlation function is zero for no
assortative mixing and positive or negative for assortative or
disassortative mixing respectively.  For the purposes of comparing
different networks, it is convenient to normalize it by dividing by its
maximal value, which it achieves on a perfectly assortative network,
i.e.,~one with $e_{jk}=q_k\delta_{jk}$.  This value is equal to the
variance $\sigma_q^2 = \sum_k k^2 q_k - \bigl[ \sum_k k q_k \bigr]^2$ of
the distribution~$q_k$, and hence the normalized correlation function is
\begin{equation}
r = {1\over\sigma_q^2} \sum_{jk} jk(e_{jk}-q_jq_k),
\label{defschi1}
\end{equation}
which is simply the Pearson correlation coefficient of the degrees at
either ends of an edge and lies in the range $-1\le r\le1$~\footnote{The
  quantity~$r$ can easily be generalized to the case of a directed network,
  where $e_{jk}$ is asymmetric and $r=\sum_{jk}
  jk(e_{jk}-q_jq_k)/(\sigma_{\rm in}\sigma_{\rm out})$, with $\sigma_{\rm
    in}$ and $\sigma_{\rm out}$ being the standard deviations of the
  remaining degrees at the in-going and out-going ends of the edge
  respectively.}.  For the practical purpose of evaluating $r$ on an
observed network, we can rewrite~\eref{defschi1} as
\begin{equation}
r = {M^{-1} \sum_i j_i k_i -
    \bigl[ M^{-1} \sum_i \half(j_i+k_i) \bigr]^2\over
    M^{-1} \sum_i \half(j_i^2+k_i^2) -
    \bigl[ M^{-1} \sum_i \half(j_i+k_i) \bigr]^2},
\label{defschi2}
\end{equation}
where $j_i,k_i$ are the degrees of the vertices at the ends of the $i$th
edge, with $i=1\ldots M$~\footnote{One can use either the total degrees or
the remaining degrees to evaluate Eq.~\eref{defschi2}---the answer is the
same either way.  Note also that we have written Eq.~\eref{defschi2} in a
form manifestly symmetric in $j_i$ and~$k_i$, so that it doesn't matter
which end of an edge is which.}.

\begin{table}[t]
\begin{tabular}{l|l|r|r}
& network                                 & $n$           & $r$       \\
\hline
\begin{rotate}{90}
\hbox{\hspace{-86pt}real-world networks}
\end{rotate}
& physics coauthorship$^{\rm a}$           & $52\,909$     & $0.363$  \\
& biology coauthorship$^{\rm a}$           & $1\,520\,251$ & $0.127$  \\
& mathematics coauthorship$^{\rm b}$       & $253\,339$    & $0.120$  \\
& film actor collaborations$^{\rm c}$      & $449\,913$    & $0.208$  \\
& company directors$^{\rm d}$              & $7\,673$      & $0.276$  \\
\cline{2-4}
& Internet$^{\rm e}$                       & $10\,697$     & $-0.189$ \\
& World-Wide Web$^{\rm f}$                 & $269\,504$    & $-0.065$ \\
& protein interactions$^{\rm g}$           & $2\,115$      & $-0.156$ \\
& neural network$^{\rm h}$                 & $307$         & $-0.163$ \\
& food web$^{\rm i}$                       & $92$          & $-0.276$ \\
\hline
\begin{rotate}{90}
\hbox{\hspace{-24pt}models}
\end{rotate}
& random graph$^{\rm u}$                   & & $0$                    \\
& Callaway~\etal$^{\rm v}$                 & & $\delta/(1+2\delta)$   \\
& Barab\'asi and Albert$^{\rm w}$          & & $0$
\end{tabular}
\caption{Size~$n$ and assortativity coefficient~$r$ for a number of
different networks: collaboration networks of (a)~scientists in physics and
biology~\cite{Newman01a}, (b)~mathematicians~\cite{GI95}, (c)~film
actors~\cite{WS98}, and (d)~businesspeople~\cite{DYB01}; (e)~connections
between autonomous systems on the Internet~\cite{Chen02}; (f)~undirected
hyperlinks between Web pages in a single domain~\cite{BA99b};
(g)~protein-protein interaction network in yeast~\cite{Jeong01};
(h)~undirected (and unweighted) synaptic connections in the neural network
of the nematode \textit{C. Elegans}~\cite{WS98}; (i)~undirected
trophic relations in the food web of Little Rock Lake,
Wisconsin~\cite{Martinez91}.  The last three lines give analytic results
for model networks in the limit of large network size: (u)~the random graph
of Erd\H{o}s and R\'enyi~\cite{Bollobas01}; (v)~the grown graph model of
Callaway~\etal~\cite{Callaway01}; (w)~the preferential attachment model of
Barab\'asi and Albert~\cite{BA99b}.}
\label{tabr}
\end{table}

In Table~\ref{tabr} we show values of~$r$ for a variety of real-world
networks.  As the table shows, of the social networks studied (the top five
entries in the table) all have significant assortative mixing, which
accords with accepted wisdom within the sociological community.  By
contrast, the technological and biological networks studied (the middle
five entries) all have disassortative mixing---high degree vertices
preferentially connect with low degree ones and \textit{vice versa}.
Various explanations for this observation suggest themselves.  In the case
of the Internet, for example, it appears that the high degree vertices
mostly represent connectivity providers---telephone companies and other
communications carriers---who typically have a large number of connections
to clients who themselves have only a single connection~\cite{Chen02}.
Thus the high-degree vertices do indeed tend to be connected to the
low-degree ones.

We have also calculated~$r$ analytically for three models of networks:
(1)~the random graph of Erd\H{o}s and R\'enyi~\cite{Bollobas01}, in which
edges are placed at random between a fixed set of vertices; (2)~the grown
graph model of Callaway~\etal~\cite{Callaway01}, in which both edges and
vertices are added at random at constant but possibly different rates, the
ratio of the rates being denoted~$\delta$; (3)~the grown graph model of
Barab\'asi and Albert~\cite{BA99b}, in which both edges and vertices are
added, and one end of each edge is added with linear preferential
attachment.

For the random graph, since edges are placed at random without regard to
vertex degree it follows trivially that $r=0$ in the limit of large graph
size.  The model of Callaway~\etal\ however, although apparently similar in
construction, gives a markedly different result.  From Eq.~(21) of
Ref.~\onlinecite{Callaway01}, $e_{jk}$~for this model satisfies the
recurrence relation
\begin{equation}
(1+4\delta) e_{jk} = 2\delta (e_{j-1,k}+e_{j,k-1}) + p_j p_k,
\end{equation}
and the degree distribution is $p_k=(2\delta)^k/(1+2\delta)^{k+1}$.
Substituting into Eq.~\eref{defschi1} and making use of
Eq.~\eref{sumrules}, we then find that $r=\delta/(1+2\delta)$.  Thus the
model shows significant assortative mixing, with a maximum value of
$r=\half$ in the limit of large~$\delta$.  This agrees with
intuition~\cite{Callaway01}: in the grown graph the older vertices have
higher degree and also tend to have higher probability of being connected
to one another, simply by virtue of being around for longer.  Thus one
would expect positive assortative mixing.

The model of Barab\'asi and Albert~\cite{BA99b} provides an interesting
counter-example to this intuition.  Although this is a grown graph model,
in which again older vertices have higher degree~\cite{AH00}, it shows no
assortative mixing at all.  Making use of Eq.~(42) of
Ref.~\onlinecite{KR01} we can show that $e_{jk}$ for the model of
Barab\'asi and Albert goes asymptotically as $1/(j^2k^2)-6/(j+k)^4$ in the
limit of large $j$ and~$k$, which implies that $r\to0$ as $(\log^2N)/N$ as
$N$ becomes large.  The model of Barab\'asi and Albert has been used as a
model of the structure of the Internet and the World-Wide Web.  Since these
networks show significant disassortative mixing however (Table~\ref{tabr}),
it is clear that the model is incomplete.  It is an interesting open
question what type of network evolution processes could explain the values
of~$r$ observed in real-world networks.

Turning now to theoretical developments, we propose a simple model of an
assortatively mixed network, which is exactly solvable for many of its
properties in the limit of large graph size.  Consider the ensemble of
graphs in which the distribution $e_{jk}$ takes a specified value.  This
defines a random graph model similar in concept to the random graphs with
specified degree sequence~\cite{BC78,MR95,NSW01}, except for the added
element of assortative mixing.

Consider a typical member of this ensemble in the limit of large graph
size, and consider a randomly chosen edge in that graph, one end of which
is attached to a vertex of degree~$j$.  We ask what the probability
distribution is of the number of other vertices reachable by following that
edge.  Let this probability distribution be generated by a generating
function $G_j(x)$, which depends in general on the degree~$j$ of the
starting vertex.  By arguments similar to those of Ref.~\onlinecite{NSW01},
we can show that $G_j(x)$ must satisfy a self-consistency condition of the
form
\begin{equation}
G_j(x) = x {\sum_k e_{jk} \bigl[G_k(x)\bigr]^k\over\sum_k e_{jk}},
\label{defsg}
\end{equation}
while the number of vertices reachable from a randomly chosen vertex is
generated by
\begin{equation}
H(x) = x p_0 + x \sum_{k=1}^\infty p_k \bigl[G_{k-1}(x)\bigr]^k.
\end{equation}
The average size of the component to which such a vertex belongs is given
by the derivative of~$H$: $\av{s} = H'(1) = 1 + \sum_k k p_k G_{k-1}'(1)$.
Differentiating Eq.~\eref{defsg} we then get
\begin{equation}
\av{s} = 1 - z\vq\cdot\vA^{-1}\cdot\vq,
\label{avs2}
\end{equation}
where $z$ is the mean degree, $\vq$ is the vector whose elements are
the~$q_k$, and $\vA$ is the asymmetric matrix with elements $A_{jk} =
ke_{jk} - q_k\delta_{jk}$.

Equation~\eref{avs2} diverges at the point at which the determinant
of~$\vA$ is zero.  This point marks the phase transition at which a giant
component forms in our graph.  By considering the behavior of
Eq.~\eref{avs2} close to the transition, where $\av{s}$ must be large and
positive in the absence of a giant component, we deduce that a giant
component exists in the network when $\det\vA>0$.  This is the appropriate
generalization for a network with assortative mixing of the criterion of
Molloy and Reed~\cite{MR95} for the existence of a giant component.

To calculate the size~$S$ of the giant component, we define $u_k$ to be the
probability that an edge connected to a vertex of remaining degree~$k$
leads to another vertex that does not belong to the giant component.  Then
\begin{equation}
S = 1 - p_0 - \sum_{k=1}^\infty p_k u_{k-1}^k,\qquad
u_j = {\sum_k e_{jk} u_k^k\over\sum_k e_{jk}}.
\label{gcs}
\end{equation}
As with most other random graph models, including the original model of
Erd\H{o}s and R\'enyi, it is usually not possible to solve for~$S$ in
closed form, but we can determine it by numerical iteration from a suitable
set of starting values for~$u_k$.

To test these results and to help form a more complete picture of the
properties of assortatively mixed networks, we have also performed computer
simulations, generating networks with given values of $e_{jk}$ and
measuring their properties directly.  Generating such networks is not
entirely trivial.  One cannot simply draw a set of degree pairs $(j_i,k_i)$
for edges~$i$ from the distribution~$e_{jk}$, since such a set would almost
certainly fail to satisfy the basic topological requirement that the number
of edges ending at vertices of degree~$k$ must be a multiple of~$k$.
Instead therefore we propose the following Monte Carlo algorithm for
generating graphs.

First, we generate a random graph with the desired degree distribution
according to the prescription given in Ref.~\onlinecite{MR95}.  Then we
apply a Metropolis dynamics to the graph in which on each step we choose at
random two edges, denoted by the vertex pairs, $(v_1,w_1)$ and $(v_2,w_2)$,
that they connect.  We measure the remaining degrees $(j_1,k_1)$ and $(j_2,
k_2)$ for these vertex pairs, and then replace the edges with two new ones
$(v_1,v_2)$ and $(w_1,w_2)$ with probability
$\min(1,(e_{j_1j_2}e_{k_1k_2})/(e_{j_1k_1}e_{j_2k_2}))$.  This
dynamics conserves the degree sequence, is ergodic on the set of graphs
having that degree sequence, and, with the choice of acceptance probability
above, satisfies detailed balance for state probabilities $\prod_i
e_{j_ik_i}$, and hence has the required edge distribution $e_{jk}$ as its
fixed point.

As an example, consider the symmetric binomial form
\begin{equation}
e_{jk} = \cN \e^{-(j+k)/\kappa} \biggl[ {j+k\choose j} p^j q^k +
                                   {j+k\choose k} p^k q^j \biggr],
\label{binomial}
\end{equation}
where $p+q=1$, $\kappa>0$, and $\cN=\half(1-\e^{-1/\kappa})$ is a
normalizing constant.  (The binomial probabilities $p$ and $q$ should not
be confused with the quantities $p_k$ and $q_k$ introduced earlier.)  This
distribution is chosen for analytic tractability, although its behavior is
also quite natural: the distribution of the sum $j+k$ of the degrees at the
ends of an edge falls off as a simple exponential, while that sum is
distributed between the two ends binomially, the parameter~$p$ controlling
the assortative mixing.  From Eq.~\eref{defschi1}, the value of~$r$ is
\begin{equation}
r = {8pq-1\over2\e^{1/\kappa}-1+2(p-q)^2},
\end{equation}
which can take both positive and negative values, passing through zero when
$p=p_0=\half-\frac14\sqrt{2}=0.1464\ldots$

\begin{figure}
\begin{center}
\resizebox{\figurewidth}{!}{\includegraphics{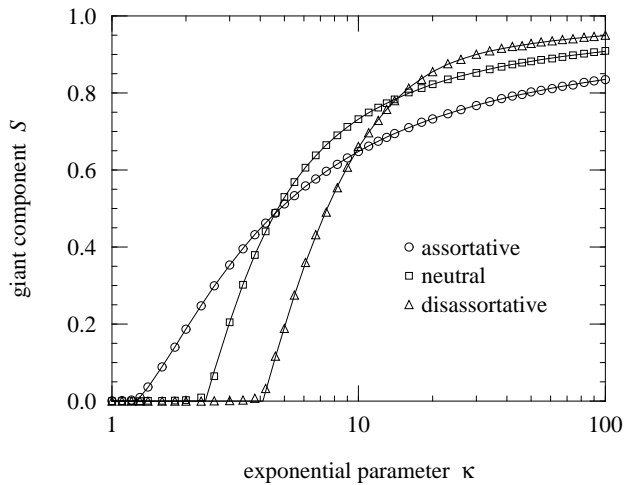}}
\end{center}
\caption{Size of the giant component as a fraction of graph size for graphs
  with the edge distribution given in Eq.~\eref{binomial}.  The points are
  simulation results for graphs of $N=100\,000$ vertices while the solid
  lines are the numerical solution of Eq.~\eref{gcs}.  Each point is an
  average over ten graphs; the resulting statistical errors are smaller
  than the symbols.  The values of~$p$ are 0.5 (circles), $p_0=0.146\ldots$
  (squares), and $0.05$ (triangles).}
\label{gc}
\end{figure}

In Fig.~\ref{gc} we show the size of the giant component for graphs of this
type as a function of the degree scale parameter~$\kappa$, from both our
numerical simulations and the exact solution above.  As the figure shows,
the two are in good agreement.  The three curves in the figure are for
$p=0.05$, where the graph is disassortative, $p=p_0$, where it is neutral
(neither assortative nor disassortative), and $p=0.5$, where it is
assortative.

As $\kappa$ becomes large we see the expected phase transition at which a
giant component forms.  There are two important points to notice about the
figure.  First, the position of the phase transition moves lower as the
graph becomes more assortative.  That is, the graph percolates more easily,
creating a giant component, if the high-degree vertices preferentially
associate with other high-degree ones.  Second, notice that, by contrast,
the size of the giant component for large~$\kappa$ is smaller in the
assortatively mixed network.

These findings are intuitively reasonable.  If the network mixes
assortatively, then the high-degree vertices will tend to stick together in
a subnetwork or core group of higher mean degree than the network as a
whole.  It is reasonable to suppose that percolation would occur earlier
within such a subnetwork.  Conversely, since percolation will be restricted
to this subnetwork, it is not surprising that the giant component has a
smaller size in this case than when the network is disassortative.  These
results could have implications, for example, for the spread of disease on
social networks~\cite{Morris93}---social networks being assortatively mixed
in many cases, as Table~\ref{tabr} shows.  The core group of an
assortatively mixed network could form a ``reservoir'' for disease,
sustaining an epidemic even in cases in which the network is not
sufficiently dense on average for the disease to persist.  On the other
hand, one would expect the disease to be restricted to a smaller segment of
the population in such cases than for diseases spreading on neutral or
disassortative networks.

Assortative mixing also has implications for questions of network
resilience, the subject of much discussion in the recent
literature~\cite{AJB00,CEBH00,CEBH01,CNSW00,Holme02}.  It has been found
that the connectivity of many networks (i.e.,~the existence of paths
between pairs of vertices) can be destroyed by the removal of just a few of
the highest degree vertices, a result that may have applications in, for
example, vaccination strategies~\cite{PV02}.  In assortatively mixed
networks, however, we find numerically that removing high-degree vertices
is a relatively inefficient strategy for destroying network connectivity,
presumably because these vertices tend to be clustered together in the core
group, so that removing them is somewhat redundant.  In a disassortative
network with a similarly sized giant component attacks on the highest
degree vertices are much more effective, these vertices being broadly
distributed over the network and presumably therefore forming links on many
paths between other vertices.  For networks of the type described by
Eq.~\eref{binomial} we find that the number of high-degree vertices that
need to be removed to destroy similarly sized giant components is greater
by a factor of about five to ten in an assortative network ($p=0.5$) than
in a disassortative one ($p=0.05$) for the typical parameter values studied
here.

These considerations paint rather a grim picture: the networks that we
might want to break up, such as the social networks that spread disease,
appear to be assortative, and therefore are resilient, at least against
simple targeted attacks such as attacks on the highest degree vertices.
And yet at the same time the networks that we would wish to protect,
including technological networks such as the Internet, appear to be
disassortative, and are hence particularly vulnerable.

To conclude, in this paper we have studied assortative mixing by degree in
networks---the tendency for high-degree vertices to associate
preferentially with other high-degree vertices.  We have defined a scalar
measure of assortative mixing and used it to show that many social networks
have significant assortative mixing, while technological and biological
networks seem to be disassortative.  We have also proposed a model of an
assortatively mixed network, which we have solved exactly using generating
function techniques, and also simulated using a Monte Carlo graph sampling
method.  Within this model we find that assortative networks percolate more
easily and that they are also more robust to removal of their highest
degree vertices, while disassortative networks percolate less easily and
are more vulnerable.  This suggests that social networks may be robust to
intervention and attack while technological networks are not.

\smallskip
The author thanks Duncan Callaway, Michelle Girvan, Cris Moore, and Martina
Morris for helpful comments, and L\'aszl\'o Barab\'asi, Jerry Davis, Jerry
Grossman, Hawoong Jeong, Neo Martinez, and Duncan Watts for providing
network data used in the calculations for Table~\ref{tabr}.  This work was
funded in part by the National Science Foundation under grant DMS--0109086.

\end{document}